\documentstyle{BSAXwork}
\def\aut{Giommi et al.}

\input epsf

\def \AAP #1 #2 {{\em Astron. Astrophys.\/} {\bf #1}, #2}
\def \AAL #1 #2 {{\em Astron. Astrophys. Lett.\/} {\bf #1}, L#2}
\def \AAR #1 #2 {{\em Astron. Astrophys. Rev.\/} {\bf #1}, #2}
\def \AAS #1 #2 {{\em Astron. Astrophys. Suppl. Ser.\/} {\bf #1}, #2}
\def \AJ #1 #2 {{\em Astron. J.\/} {\bf #1}, #2}
\def \ANNREV #1 #2 {{\em Ann. Rev. Astron. Astrophys.\/} {\bf #1}, #2}
\def \APJ #1 #2 {{\em Astrophys. J.\/} {\bf #1}, #2}
\def \APJL #1 #2 {{\em Astrophys. J. Lett.\/} {\bf #1}, L#2}
\def \APJS #1 #2 {{\em Astrophys. J. Suppl.\/} {\bf #1}, #2}
\def \APSS #1 #2 {{\em Astrophys. Space Sci.\/} {\bf #1}, #2}
\def \ASR #1 #2 {{\em Adv. Space Res.\/} {\bf #1}, #2}
\def \BAIC #1 #2 {{\em Bull. Astron. Inst. Czechosl.\/} {\bf #1}, #2}
\def \JSQRT #1 #2 {{\em J. Quant. Spectrosc. Radiat. Transfer\/} {\bf #1}, #2}
\def \MN #1 #2 {{\em Mon. Not. R. Astr. Soc.\/} {\bf #1}, #2}
\def \MNs #1 {{\em Mon. Not. R. Astr. Soc.\/} {\bf #1}}
\def \MEM #1 #2 {{\em Mem. R. Astr. Soc.\/} {\bf #1}, #2}
\def \PLR #1 #2 {{\em Phys. Lett. Rev.\/} {\bf #1}, #2}
\def \PASJ #1 #2 {{\em Publ. Astron. Soc. Japan\/} {\bf #1}, #2}
\def \PASP #1 #2 {{\em Publ. Astr. Soc. Pacific\/} {\bf #1}, #2}
\def \NAT #1 #2 {{\em Nature\/} {\bf #1}, #2}
\def \SAIT #1 #2 {{\em Mem.\ Soc.\ Astron.\ It.\/} {\bf #1}, #2}
\def \MESS #1 #2 {{\em The Messenger\/} {\bf #1}, #2}
\def \ASTRNACH #1 #2 {{\em Astron. Nach.\/} {\bf #1}, #2}

\def\sax{{\em Beppo}SAX}
\def\aro{{$\alpha_{\rm ro}$~}}
\def\aox{{$\alpha_{\rm ox}$~}}
\def\arx{{$\alpha_{\rm rx}$~}}
\def\ergs{{erg~cm$^{-2}$s$^{-1}$~}}
\def\ergj{{erg~cm$^{-2}$s$^{-1}$Jy$^{-1}$~}}
\def\vovm{{$V/V_{\rm m}$}~}
\def\veova{{$V_{\rm e}/V_{\rm a}$}~}
\def\fxfr{$f_{\rm x}/f_{\rm r}$~}
\def\fr{$f_{\rm r}$~}
\def\fx{$f_{\rm x}$~}
\def\nupeak{$\nu_{peak}$~}

\def\vovaave{{$\langle V_{\rm e}/V_{\rm a} \rangle$}~}
\def\gsim{\ \raise -2.truept\hbox{\rlap{\hbox{$\sim$}}\raise5.truept\hbox{$>$}\
}}

\epsfverbosetrue
\begin{opening}

\title{Parameter Correlations and Cosmological Properties of BL Lac Objects} 
\author{P.Giommi $^{1,2}$ P. Padovani $^{3,4}$ M. Perri $^{1,5}$ H. Landt $^{3}$ and E. Perlman $^{6}$}
\institute{$^1$ASI Science Data Center, c/o ESA-ESRIN, Frascati, Italy\\
$^2$Agenzia Spaziale Italiana, Italy \\
$^3$Space Telescope Science Institute, 3700 San Martin Drive, Baltimore MD 21218, USA\\
$^4$On assignement from the Space Telescope Operations Division of ESA\\
$^5$Dip. di Fisica, Universit\`a ``La Sapienza'', P.zle A. Moro 2,
I-00185 Roma, Italy\\
$^6$Joint Center for Astrophysics, University of Maryland, Baltimore, MD 21250, USA \\
}
\date{} 
\end{opening}

\begin{document}

\oddpagefooter{}{}{} 
\evenpagefooter{}{}{} 
\medskip  

\begin{abstract} 
Using three complete, radio flux limited, blazar samples
we compare the LogN-LogS and the preliminary radio luminosity function of the general 
population of BL Lacs to those of the subclass of high energy synchrotron 
peaked (HBL) BL Lacs. We also examine recent results on the cosmological 
evolution in different samples of BL Lacs and 
we investigate the controversial issue of the correlation between the 
synchrotron peak frequency and radio luminosity in BL Lacertae objects. 
We find that the fraction of HBL objects is approximately the same 
at all observed radio fluxes and luminosities implying that there cannot be any 
strong correlation between the position of the synchrotron peak and radio luminosity. 
The amount of cosmological evolution in BL Lacs is confirmed to be low and negative 
at low radio fluxes, although the large number of objects  without
redshift prevents a precise estimation. At high radio fluxes the amount of cosmological 
evolution is zero or slightly positive but this could be induced by a possible contamination 
with Flat Spectrum Radio Quasars.
 
\end{abstract}

\medskip

\section{Introduction}

The blazar class includes BL Lacertae type objects (BL Lacs) and Flat Spectrum 
Radio Quasars (FSRQs), both characterized by a strong and highly variable non-thermal 
continuum across the entire electromagnetic spectrum. 
The often extreme and peculiar observational properties of these sources  (e.g. Urry \& Padovani 1995) 
are thought to be the signature of physical phenomena seen in rare conditions such 
as the electromagnetic emission arising from a jet of material moving toward the observer 
at relativistic speeds observed at small viewing angles. For these reasons blazars 
are intensively studied at all frequencies from radio to high energy gamma rays. However, 
their low surface density and the intrinsic difficulties in discovering them
at optical frequencies where BL Lacs do not display strong emission 
lines nor show any Blue Bump, make blazar research a difficult business. 
So difficult, in fact, that it took nearly 20 years to build the first statistically 
complete samples such as the 1Jy and 2Jy samples at 5GHz 
(Stickel et al. 1991, Wall \& Peacock 1985) and the X-ray flux-limited samples 
derived from the {\it Einstein} EMSS and Slew surveys (Stocke et al. 1991, 
Rector et al. 2000, Perlman et al. 1996a).  
Most of the experimental results on which our present understanding of blazars is based 
come form the study of these early samples which include 30-50 objects, 
the brightest and most luminous in their selection band.

Despite the early difficulties many new surveys,
based on classical or more recent multi-frequency selection techniques, are now becoming 
available. These surveys, however are characterized by different degrees of completeness 
and by a range of sensitivity limits in one or more energy bands; see Padovani (2002) for 
a recent review.  

Using three recent, radio flux-limited and nearly complete surveys, which cover 
a portion of the blazar parameter space much wider than earlier samples, 
we determine some cosmological properties of BL Lacs such as the radio LogN-LogS, 
the radio luminosity function and its cosmological evolution and we address 
the controversial issue of the experimental basis of the often quoted correlation between 
\nupeak and radio luminosity (Fossati et al. 1998). 
Throughout this paper we assume $H_0=50~$Km~s$^{-1}$Mpc$^{-1}$ and $q_0=0$.

\section{Blazars surveys}

The statistical estimation of a parameter value, or the verification 
of the existence of a correlation between two or more parameters, in a population of 
sources requires the use of well defined, completely identified, possibly 
sizable, unbiased samples. A sample that includes {\em all} objects that are detectable 
above its statistical limits is usually referred to as a {\em complete} sample.
The 'biasness' of a statistical sample depends on the parameter(s) that are 
estimated through it. A sample is unbiased with respect to a given parameter if,
above the survey statistical limits, the selection method used neither favors, 
nor selects against, objects with particular values of the parameter(s) considered.
Surveys of X-ray serendipitous sources are particularly delicate from this viewpoint since they 
tend to exclude bright sources that have been observed as targets in the main 
observation program and therefore have much less probability to be rediscovered as
serendipitous sources. 
For instance the samples based on ROSAT serendipitous sources like the DXRBS
and the REX must suffer of some incompleteness at bright radio (\fr $\gsim$ 1 Jy) 
and at bright X-ray (\fx $>$ a few $10^{-12}$ \ergs) fluxes since the ROSAT PSPC 
pointing program included nearly all the BL Lacs from the 1Jy sample and a large faction of 
the BL Lacs discovered in the {\em Einstein} EMSS and Slew Surveys. The exact amount 
depends on the details of the survey.

\setcounter{figure}{0}
\begin{figure}[!h]
\vspace*{-1.5cm}
\centering
\epsfysize=9.0cm\epsfbox{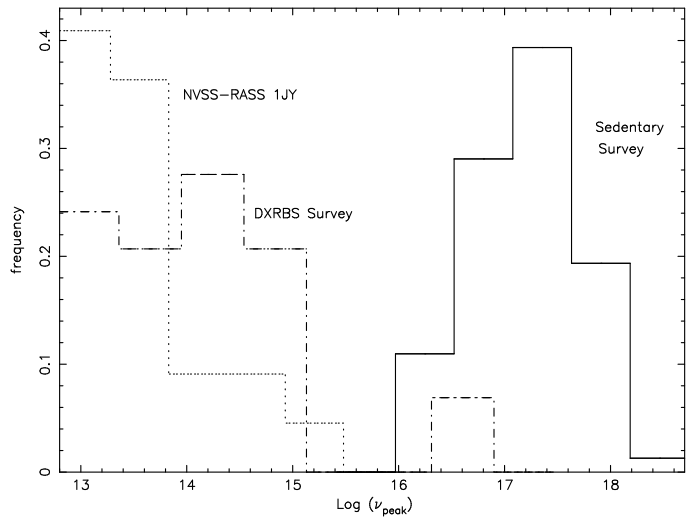}
\caption[h]{The BL Lac \nupeak distribution in the Sedentary survey (solid histogram),
the NVSS-RASS 1Jy survey (dotted histogram) and the DXRBS survey 
(dash-dotted histogram). The radically different \nupeak distribution of the 
Sedentary survey is the result of selection and shows how this survey is efficient 
in finding rare objects that are barely represented in other radio flux limited surveys.}
\label{nupeaks}
\end{figure}


We make use of three radio flux limited blazar surveys, namely the DXRBS survey 
(Perlman et al. 1998 and Landt et al. 2001), the NVSS-RASS 1Jy sample (Giommi et al. 2002)
and the Sedentary survey of extreme HBL BL Lacs (Giommi, Menna \& Padovani 1999),
whose spectroscopic identification rate has reached a sufficiently high level of completion ($\gsim 85\%$) 
and are expected to be unbiased with respect to the quantities that we want
to estimate. These large samples allow us to draw firm, albeit preliminary, conclusions with respect 
to the LogN-LogS, the luminosity function and the cosmological evolution of BL Lacs.
In addition, the comparison of the NVSS-RASS 1Jy and the DXRBS surveys, which select 
objects regardless of their \nupeak value, to the Sedentary survey, which instead 
only includes high \nupeak sources, allow us to verify the experimental bases of 
the radio luminosity \nupeak correlation reported by Fossati et al. (1998).



In the following we briefly describe the main characteristics of these samples 
referring the reader to the original papers for more details.

\subsection{The Deep X-ray Radio Blazars Survey, DXRBS}

The DXRBS survey efficiently selects blazars by means of a spatial cross-correlation of
the ROSAT WGA catalog of soft X-ray sources with several radio catalogs (GB6, 
PMN at 5~GHz and NORTH20CM at 1.4~GHz) and restricting the sample to objects with flat radio 
spectrum ($\alpha_r <0.7, S_{\nu} \propto \nu^{-\alpha_r}$); see Perlman et al. 1998 and 
Landt et al. 2001 for details.
All optical candidates are observed for spectroscopic identification down
to the 24th magnitude.
The flux limits and the present status of the DXRBS is briefly summarized below.

\noindent  - X-ray flux limits between $\approx 10^{-14}$ and $\approx 10^{-12}$ \ergs \\
 - Radio flux limits between 50 and 100 mJy at 5 GHz\\
 - optical identification program is 95\% complete\\
 - 233 confirmed blazars, 199 FSRQ + 44 BL Lacs\\
 - about 30\% of the BL Lacs have no redshift \\

This survey is complete above its radio flux limits with a sky coverage that depends mainly on X-ray 
and only weakly on radio flux. A small number of bright radio (or X-ray) blazars are missing because they were 
targets of pointed PSPC observations.

\subsection{The Sedentary multi-frequency survey.}

The Sedentary multi-frequency survey is a radio flux limited sample of extreme HBL BL Lacs 
(that is objects with very high \fxfr  or, equivalently, very high energy synchrotron 
peak frequency). It is based on the cross-correlation between the NVSS catalog
of 20~cm radio sources and the ROSAT all sky survey (RASS) bright sample 
with the restriction that the \aro and \aox of these candidates are within
the "HBL zone", that is the small area of the \aox -- \aro plane that is almost 
exclusively populated by extreme HBL (\arx $<$ 0.55, or \nupeak $\gsim 10^{16}$Hz; see 
Giommi et al. 1999 for details). 

The current status of the survey is presented in Perri et al. 2002; here we summarize 
the main properties.

\noindent
- 155 sources: the largest complete sample of BL Lacs \\
- radio flux limit 3.5 mJy at 1.4 GHz\\
- $\sim 83$\% spectroscopically identified \\
- $\approx 40 \%$ of the objects have no redshift \\

\subsection{The NVSS-RASS : a new 1Jy Blazar survey at 1.4 GHz}

Sample similar to the "classical" 1Jy sample at 5GHz of Stickel et al. (1991) based on
the cross-correlation between NVSS (1.4 GHz) radio sources with flux larger that 1 Jansky 
and the ROSAT All Sky Survey (RASS, bright and faint samples). 
From the numerous X-ray observations of BL Lacs over the past 25 years it has 
clearly emerged that all objects of this type emit X-rays with an \fxfr ratio that 
ranges from $\approx 2\times 10^{-13}$ to well over $10^{-10}$ \ergj (e.g. Padovani 2002). 
Since the RASS full sample has a range of limiting sensitivities of $\sim 5\times10^{-14}$ to 
$\sim 5\times10^{-13}$ all sources brighter than
1 Jy in the NVSS should be detected in the RASS. Therefore the sample defined above 
should be complete provided that the RASS sky coverage is taken in to account.
Currently the survey includes 

\noindent
- 226 sources, 95\% of them identified from the literature. \\
- 165 Quasars (flat + steep radio spectrum) \\
- 24 BL Lacs \\
- 19 radio galaxies\\
- 3 seyfert galaxies\\
- 12 still unidentified\\

\par 
Figure \ref{nupeaks} shows the distribution of the synchrotron peak energies (\nupeak) for 
the three surveys, estimated  by comparing a SSC model to the SED  
of all the sources in the samples (see Giommi et al., these proceedings for an application of 
this method to the sample of blazars observed with \sax). Note that the \nupeak 
distribution of the DXRBS and the NVSS-RASS 1Jy overlap substantially, (these being radio 
surveys that do not select any \nupeak values) while the Sedentary survey (by construction) 
includes all objects with values of \nupeak high enough to satisfy the survey requirement 
of \arx $<$ 0.55.

\par
\section{Cosmological properties of BL Lacs}

In this paragraph we use the complete and radio flux limited samples described above to derive 
the cosmological properties of BL Lacs in general and those of the subclass of high  \nupeak sources.

\subsection{The radio LogN-LogS}

The radio LogN-LogS of the surveys considered here are plotted in Figure \ref{logns} together 
with that derived from the "classical" 1Jy sample of Stickel et al. (1991), after converting 
the 5GHz fluxes of this last survey into the 1.4 GHz band (assuming an average radio spectral
slope of 0.3).
From this figure we see that the number density of 
objects in the Sedentary survey (extreme HBL) is about 1/30th of the full population of BL Lacs and 
that this fraction remains roughly constant within the flux range where the Surveys' sensitivities 
overlap. At fluxes below 50 mJy the LogN-LogS of high \nupeak sources flattens considerably so that,
unless the LogN-logS of all BL Lacs flattens dramatically, the fraction of high \nupeak sources
remains very low at all fluxes.  
Note that above a few hundred mJy the number density 
of high \nupeak "Sedentary" sources becomes so low (a few $10^{-5} $deg$^{-2}$) that even all sky surveys 
(covering $\approx 2\times10^{4}~$deg$^{2}$ like the 1Jy surveys) are expected to include less than one 
such objects.

\setcounter{figure}{1}
\begin{figure}[!h]
\vspace*{-2.5cm}
\centering
\epsfysize=10.0cm\epsfbox{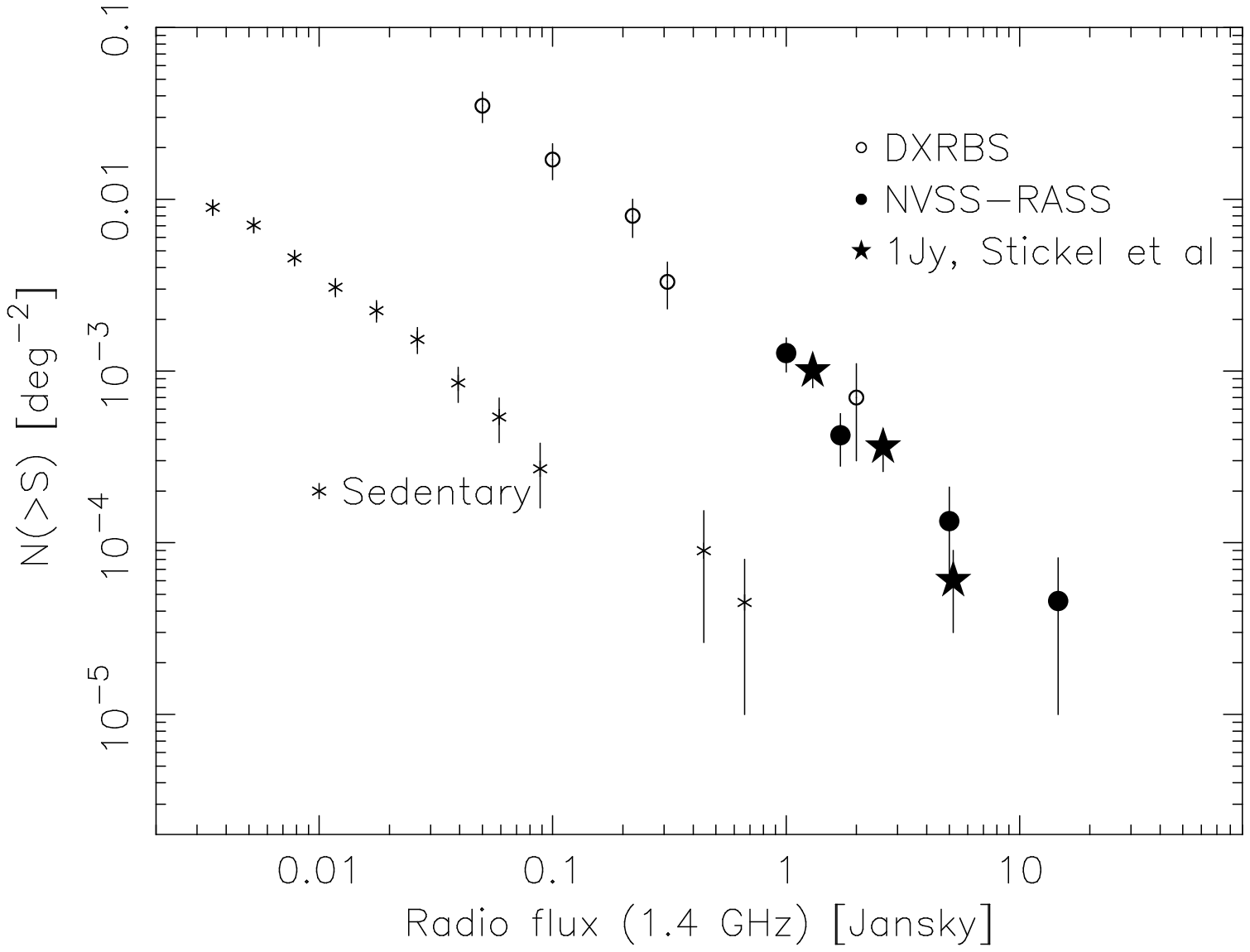}
\caption[h]{The radio LogN-LogS of BL Lacs as estimated from surveys including all
types of BL Lacs (1Jy Stickel et al. 1991, Giommi et al. 2002, and DXRBS)
and from the Sedentary survey which only includes
extreme HBL BL Lacs. The LogN-LogS of the two populations are approximately parallel
implying that the fraction of HBL sources is roughly constant at all fluxes as suggested by 
Giommi \& Padovani (1995), and Padovani \& Giommi (1996) and is not consistent
with a strong correlation between \nupeak and luminosity of Fossati et al. 1998,  which 
instead requires that the fraction of HBLs is a strong function of radio flux. }
\label{logns}
\end{figure}

\subsection{The radio luminosity function}

Figure \ref{lf} shows the radio luminosity function of all BL Lac types (LBL +HBL) estimated 
from the DXRBS and the NVSS-RASS 1Jy and 
of the subset of extreme HBL sources derived from the Sedentary survey. 
Given the high completeness level of all these surveys we do not expect that these
preliminary LF can change much when full completion is achieved. 

Since about 40\% of the BL Lacs in the Sedentary survey do not have measured redshift
we must make some assumptions to estimated the LF. We considered two cases 1) all 
objects with no redshift were assigned a value of 0.25 which is roughly equal to the 
average redshift of the remaining BL Lacs in the survey, and 2) the redshift was 
assumed to be that for which the expected optical nonthermal luminosity, estimated from  
\aro would be five times that of a typical host galaxy (give value). The LFs that
we obtained under the two hypothesis are shown as small points and filled circles
respectively. 

From Figure \ref{lf} we see that the LF of extreme HBL objects is roughly parallel to that
of the full population of BL Lacs implying that the fraction of high \nupeak objects 
is approximately the same at all luminosities.   

\setcounter{figure}{2}
\begin{figure}[!h]
\vspace*{-1.5cm}
\centering
\epsfysize=10.0cm\epsfbox{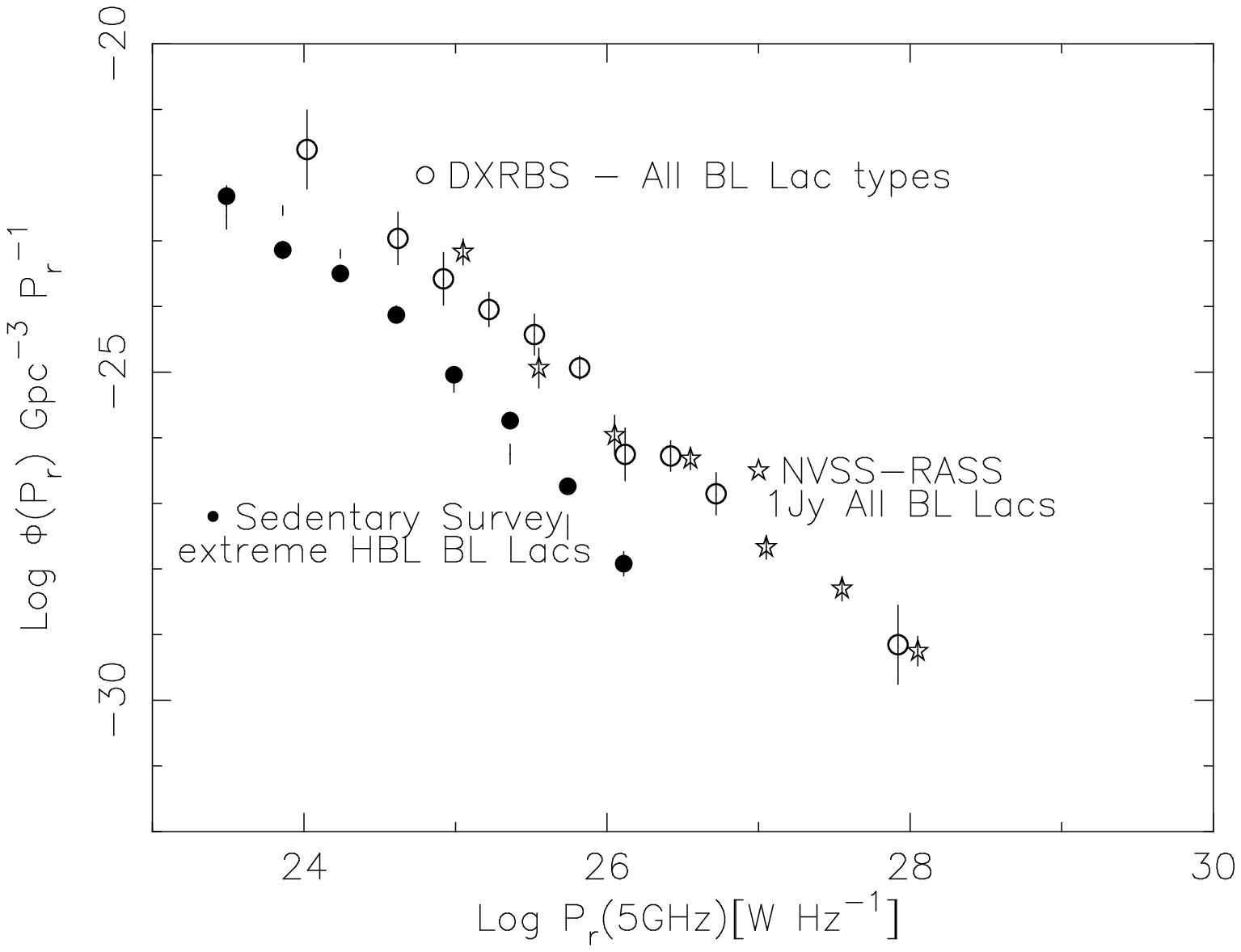}
\caption[h]{The preliminary composite radio luminosity function of all BL Lac types,
from the NVSS-RASS 1Jy surveys of Giommi et al. (2002) and Padovani (2002)
is compared to the preliminary luminosity function of the extreme HBL Lacs of the
Sedentary survey. Note that the LF of HBL Lacs is essentially parallel to that of 
the full population of BL Lac implying that the fraction of BL Lacs with very high
\nupeak value is approximately the same at all luminosities.}
\label{lf}
\end{figure}

\subsection{Cosmological evolution}

The cosmological evolution of BL Lacs is long known to be peculiar (Stocke et al. 1982, 
Maccacaro et al. 1984, Morris et al. 1991, Bade et al. 1998, Rector et al. 2000, 
Giommi et al. 1999, Caccianiga et al. 2002) and markedly lower than that of any other type of AGN. 
However, a number of studies seem to indicate that the amount of
cosmological evolution may depend on flux limits and on the type of BL Lac considered, 
with HBL sources being those with the lowest evolution.  
The best way to measure cosmological evolution in a population of cosmic sources
is to compare the luminosity function at different epochs, or equivalently, in 
different redshift shells. This, however, requires very large complete samples which are 
not easy to get. Although the Sedentary survey is large enough to allow this approach, 
the present status of the identification program (83\%) and the large fraction 
of objects without a measured (or measurable) redshift, make this approach still premature. 
For the purposes of this work we measured the amount of cosmological evolution 
through the \veova test (Avni \& Bachall 1980), an 
extension of the original \vovm method (Schmidt 1968) to surveys with more than
one flux limit. This test measures the presence of cosmological evolution as a deviation
from a uniform distribution of \veova values (and from its average value \vovaave of 0.5 ) which is
expected in a population of non evolving sources.
To assign a redshift to the sources with no measured value we have followed three 
approaches: 1) z was set equal to
the average value of the subsample with measured redshifts (z=0.25);
2) we have used a Monte Carlo method to assign redshift values assuming that the z 
distribution is the same as in the subsample of BL Lacs with measured redshifts;
3) we have applied the Monte Carlo method only to the sources that are still unidentified 
and we have derived redshift upper limits for the confirmed BL Lacs without redshift estimation 
assuming that they are so because their optical spectrum is featureless as a consequence of 
the fact that their optical Synchrotron Self Compton (SSC) power is at least five times that 
of an average BL Lac host galaxy (e.g. Landt et al. 2002).

We have obtained \vovaave = $0.41\pm0.02$, \vovaave = $0.42\pm0.02$ and \vovaave = $0.44\pm0.02$
for case 1, 2 and 3 respectively. In all cases \vovaave is less than 0.5 by three sigma or more.
Table \ref{cosmoother} compares the \vovaave results of the Sedentary survey with those of other 
recent surveys.
Low values of \vovaave are found in the Sedentary survey and in X-ray 
surveys like the EMSS and RASS-HBX, which include a large fraction of HBLs. 
A somewhat contraddictory result seems to be that of Caccianiga et al. (2002) who find the
\vovaave of the XB-REX sample to be consistent with 0.5, at odds with the measurements of 
lower values in 
the EMSS and RASS-HBX surveys which have similar X-ray flux limits. However we note that 
the XB-REX result is only two sigma 
away from the other measurements and that the REX survey must have missed a few bright (hence 
low \veova) HBLs of the {\it Einstein} Slew and EMSS surveys since a large fraction of these 
objects were pointed as part of the main ROSAT observation program.

\begin{table}[h]
\caption{Comparison of \vovaave results in different surveys}
\begin{center}
\begin{tabular}{|l|c|c|c|}
\hline 
Survey & \vovaave & Band & Flux limit \\
\hline
1Jy survey (Stickel et al. 1991) &$0.61\pm0.05$& Radio-5GHz &1 Jansky \\
NVSS-RASS 1Jy & $0.61\pm0.06$ & Radio-1.4GHz &1 Jansky \\
DXRBS  & $0.56\pm0.05$ & Radio-5GHz & 50 mJy \\
Sedentary survey & $0.41\pm0.02$ & Radio-1.4GHz & 3.5 mJy \\
{\it Einstein} EMSS & $0.43\pm0.05$& Soft X-ray & $2\times 10^{-13}$ \ergs \\
XB-REX & $0.48\pm0.04$ & Soft X-ray & $4\times 10^{-13}$ \ergs \\
RASS HBX (Bade et al. 1998) & $0.41\pm0.05$ & Soft X-ray & $8\times 10^{-13}$ \ergs \\ 
\hline
\end{tabular}
\label{cosmoother}
\end{center}
\end{table}

\section{The \nupeak-radio luminosity correlation}

In this section we address the issue of the correlation between \nupeak and radio luminosity
that has been noticed comparing radio selected and X-ray selected samples of blazars
by Fossati et al. (1998) and is considered as the experimental basis of unified scenarios
for blazars (Ghisellini et al. 1998, Costamante et al. 2001, Maraschi 2002, these proceedings). 

Figure \ref{fossati} (adapted from Costamante et al. 2001) shows the original data used by 
Fossati et al. (1998) to point out the correlation between \nupeak and radio luminosity.
Fossati et al.  compared powerful radio blazars from the 1Jy
sample of Stickel et al. (1991) and the 2Jy sample FSRQ with less radio powerful X-ray
discovered BL Lacs from the Einstein IPC medium Sensitivity and Slew Surveys, which  
at that time were the only complete samples available. 
It must be noted, however that although these samples were complete above their respective radio 
and X-ray flux limits, they included sources with very different radio fluxes, and not all radio
luminosities represented in one survey were above the flux limit in the other sample. 
Perlman et al. (2001) showed that when more recent surveys are considered the correlation 
does not appear as straightforward as in Figure \ref{fossati}.

Figure \ref{fossati_aff} shows the radio luminosity - \nupeak plot extended to BL Lacs discovered in the
deeper and more homogeneous DXRBS and Sedentary surveys.
The original data (for BL Lacs only) still appear in the plot as small circles.
It is readily apparent that the lower left corner of the plot 
(low power - low \nupeak sources), which was empty in Figure \ref{fossati_aff} is now filled  
by most of the DXRBS BL Lacs. 
The upper left corner, instead remains empty testifying that high~radio~power-high~\nupeak BL Lacs 
are not detected. However, this is probably due to
the fact that high radio power high-\nupeak sources very easily outshine the optical emission
of their host galaxy thus hampering any redshift (and luminosity) measurement. Figure \ref{gal_lbl_hbl} shows
the SED of two BL Lacs of equal radio luminosity but with synchrotron peaks 
located at low (\nupeak $\approx 10^{13}$ Hz) and at high energy (\nupeak $\approx 
10^{17}$ Hz) together with the optical emission of a typical giant elliptical 
host galaxy. It is immediately clear that the type of observed optical emission (thermal vs non-thermal),
on the basis of which an identification is made,
strongly depends on the position of \nupeak (see also Landt et al. 2002 
and Landt, these proceedings). In fact, according to the often used calcium break
dilution rule (Stocke et al. 1991, Browne \& Marcha 1993, Landt et al. 2002) the BL Lac 
with low \nupeak in Figure \ref{gal_lbl_hbl} would be classified as a radio galaxy, while the 
high energy peaked object would be classified as a featureless BL Lac with no redshift (and 
consequently no estimate of its luminosity). Since the optical luminosity of 
BL Lac host galaxies display a very small dispersion (Wurtz et al. 1997, Urry et al. 2000) the 
degree of optical light dilution can be as much as 100 times higher in HBLs than in LBL objects.\\
The dotted and dashed-dotted lines in Figure \ref{fossati_aff} represent the loci where the 
non-thermal emission from a BL Lac object is 5 times that of the host galaxy, and therefore it 
outshines any galactic component,
for two assumptions of the radio spectral index ($\alpha=0.15$ and $\alpha=0.3$). Any object 
located to the right of these lines would appear as featureless, unless 
emission lines were also present. Objects of this type cannot be plotted and the upper right
part of the plot remains empty. Several high luminosity-high \nupeak BL Lacs may be present in 
the Sedentary survey where about 40 \% of the objects have no redshift. The radio luminosity of 
these sources are plotted as lower limits (calculated as in section 3.3) in Figure \ref{fossati_aff}. Note that several of the 
1Jy BL Lacs are on the right side of the dotted line, well into the "lined objects" part of the 
diagram. That is because some 1Jy BL Lacs indeed show emission lines
(Bade et al. 1998, Rector \& Stocke 2001) and could be Flat Spectrum Radio Quasars whose
strong non-thermal emission dilutes the line strengths and lowers their equivalent width 
close to the 5 \AA ~limit. This contamination could easily affect the amount of cosmological 
evolution in bright radio flux limited samples (see also Perlman et al. 1996b). 
 
\par
\setcounter{figure}{3}
\begin{figure}[!h]
\vspace*{-1.5cm}
\centering
\epsfxsize=7.0cm\epsfbox{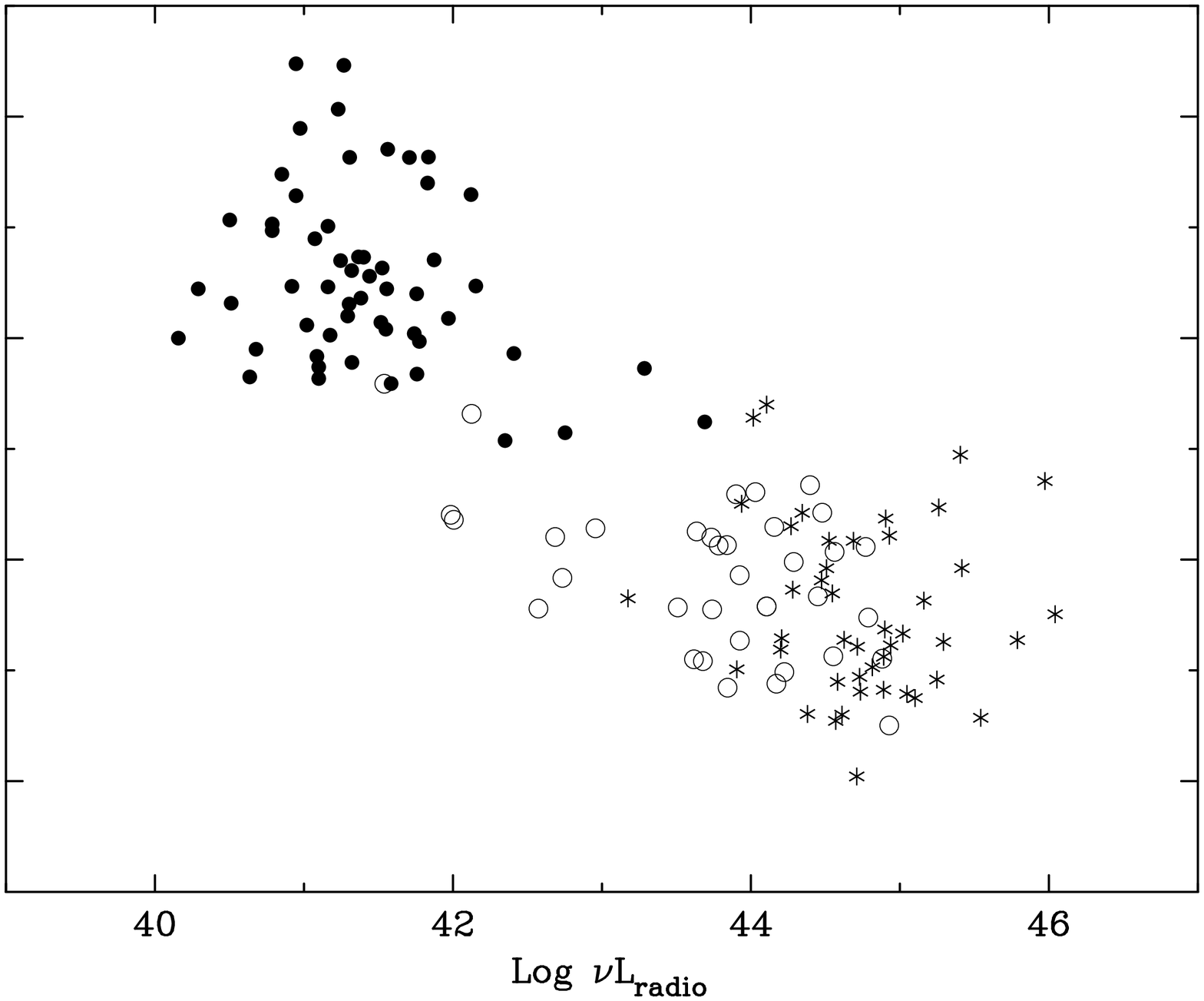} 
\caption[h]{The correlation first described by Fossati et al. 1998 as adapted from 
Costamante et al. 2001. BL Lacs from radio surveys are plotted as open circles, 
BL Lacs from X-ray surveys are shown as filled circles, and FSRQ are represented by stars.}
\label{fossati}
\end{figure}
\setcounter{figure}{4}
\begin{figure}[!h] 
\vspace*{-1.5cm}
\centering
\epsfxsize=7.0cm\epsfbox{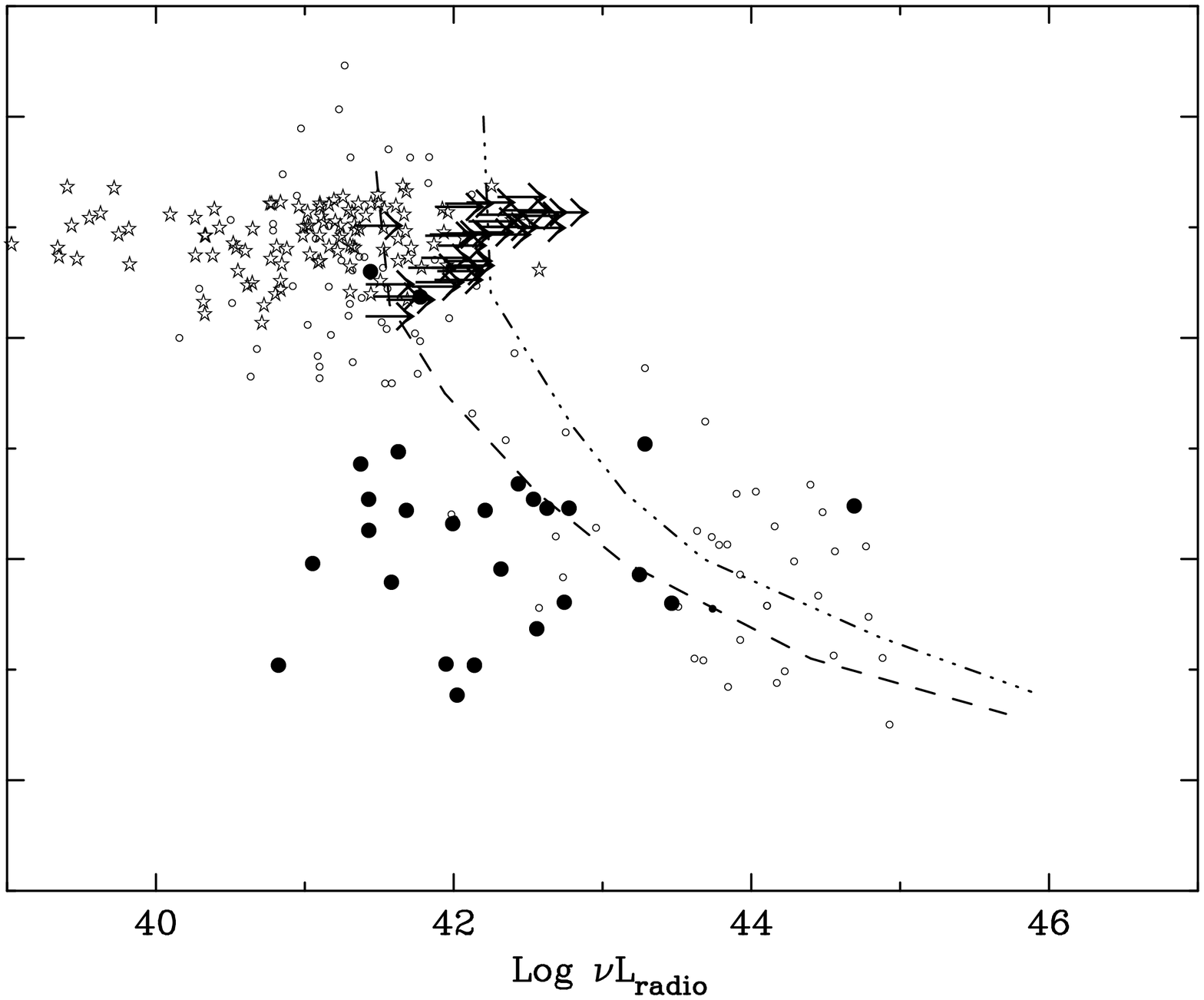}
\caption[h]{The \nupeak - radio-luminosity plane for BL Lacs from the DXRBS (filled circles) 
and from the Sedentary survey (stars for objects with measured redshift and right arrows for 
luminosity limits for objects with no redshift, see text for details). The points 
used in Figure \ref{fossati} (BL Lacs only) are also plotted as small circles. The dotted 
and dashed lines represent the the loci where the SSC emission from a BL Lac starts to be bright enough 
to outshine the host galaxy for two cases of the radio spectral index 
($\alpha=0.15$ and $\alpha=0.3$). 
The area to the right of these lines cannot include 
any object unless emission lines are also present.}
\label{fossati_aff}
\end{figure}


\section{Summary and Conclusion}

Using three new radio flux limited samples that cover a large portion of the 
blazar parameter space we have derived the cosmological properties of 
BL Lac objects and we have investigated the existence of the correlation 
between \nupeak and radio luminosity in BL Lac objects. 
We have shown that
\begin{itemize}

\item The radio LogN-LogS of extreme HBL objects remains approximately parallel to that 
of all types of BL Lacs down to 50 mJy and flattens significantly at fainter fluxes. 

\item The radio luminosity function of HBL sources also remains approximately 
parallel to that of the general population of BL Lacs
although the large number of objects without redshift leaves some uncertainties. There is 
no evidence for a strong increase in the fraction of HBL sources as would be
expected if \nupeak is strongly correlated with radio luminosity.

\item The correlation found by Fossati et al. (1998) might have been induced 
by the comparison of samples with widely different radio 
flux limits and by the fact that high luminosity HBLs were under-represented in early samples
since a) their space density at high radio fluxes is so low that hardly any object of this
type is expected in a shallow radio survey, and b) 
their flat synchrotron spectrum, which extends to very high energies, easily outshines the 
optical emission of the host galaxy thus preventing 
the measurement of their redshift and hence of their luminosity.

\item The amount of cosmological evolution in BL Lacs is confirmed to be 
low and probably negative. This is mainly confined to samples of low 
radio flux and high \nupeak objects, although some contamination by FSRQ in 
radio flux limited surveys may have artificially increased the value of \vovaave .

\item Although at the moment we do not have direct evidence of the existence of high radio power  
HBL sources, such evidence could be obtained detecting high redshift absorption
lines, due to intervening intergalactic material, in high quality spectra  
of the many featureless HBLs in the Sedentary survey. 
\end{itemize}

\setcounter{figure}{5}
\begin{figure}[!ht]
\vspace*{-2.9cm}
\centering
\epsfysize=8.0cm\epsfbox{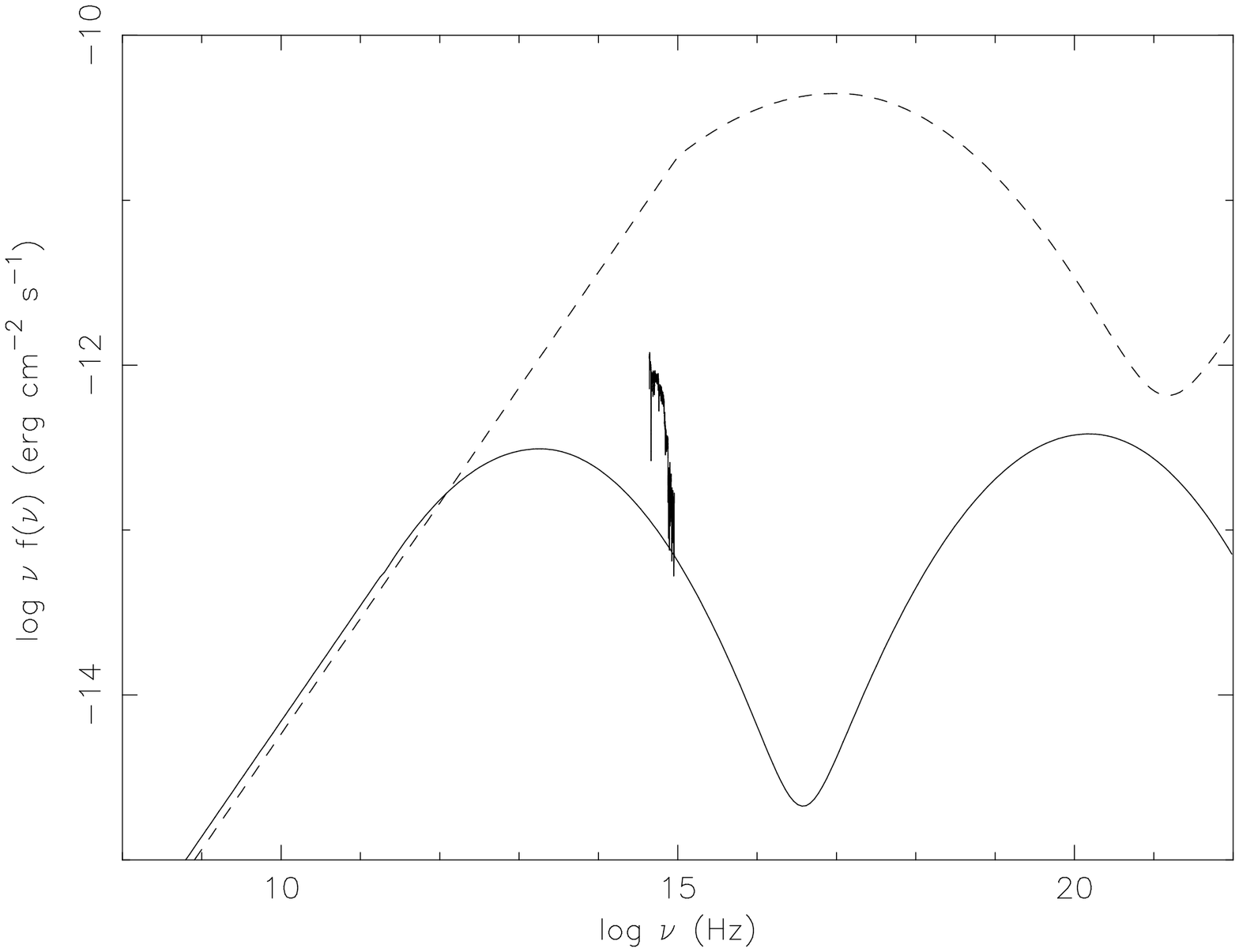}
\caption[h]{The optical emission from a typical (giant elliptical) BL Lac 
host galaxy is compared to the broad-band Synchro-Self-Compton emission of two 
hypothetical LBL and HBL sources of equal radio luminosity. While the 
emission of the HBL object outshines that of the host galaxy by a large factor, 
the optical non-thermal emission of the LBL is a small fraction 
of that of the galactic emission. In this case the HBL would appear as a 
featureless object for which no redshift would be measurable, while 
the LBL would be classified as a radio galaxy.}
\label{gal_lbl_hbl}
\end{figure}
\acknowledgements
EP acknowledges support from NASA grants NAG5-9995 and NAG5-10109 (ADP) and
NAG5-9997 (LTSA).


\end{document}